\documentclass[prl,aps,twocolumn,10pt,showpacs,nofootinbib]{revtex4-2}
\usepackage[utf8]{inputenc}
\usepackage[T1]{fontenc}
\usepackage{amsmath,amsthm}
\usepackage{amsfonts}
\usepackage{amsmath}
\usepackage{amssymb}
\usepackage{mathcmd}
\usepackage{graphicx}
\usepackage[normalem]{ulem}
\usepackage{braket}
\usepackage{bm}
\usepackage[dvipsnames]{xcolor}
\usepackage{comment}
\usepackage{tikz}
\usepackage{titlesec}
\usepackage{etoolbox}
\usetikzlibrary{decorations.pathmorphing,decorations.markings,arrows.meta,positioning,calc}
\usepackage{rotating}
\usepackage[pass]{geometry}
\usepackage[resetlabels]{multibib}
\newcites{SM}{Supplemental References} 
\makeatletter
\AtBeginDocument{%
  \@ifundefined{r@LastBibItem}%
    {\expandafter\gdef\csname r@LastBibItem\endcsname{{0}{0}}}{}%
}
\makeatother
\tikzset{
  prop R/.style={
    thick, draw=black,
    decoration={markings, mark=at position 0.55 with {\arrow{Latex[length=2.2mm]}}},
    postaction={decorate}
  },
  prop A/.style={
    thick, draw=black, dashed,
    decoration={markings, mark=at position 0.55 with {\arrow{Latex[length=2.2mm]}}},
    postaction={decorate}
  },
  prop K/.style={
    thick, draw=black,
    decorate, decoration={snake, amplitude=0.6mm, segment length=2mm}
  },
  ext r/.style={thick, draw=black},
  ext a/.style={thick, draw=black, dash pattern=on 1pt off 1.2pt},
  Vrra/.style={circle, fill=black, inner sep=1.4pt},
  Vaaa/.style={circle, draw=black, fill=white, inner sep=1.6pt, line width=0.6pt},
}

\usepackage[colorlinks=true, linkcolor=purple, citecolor=blue]{hyperref}
\newcommand{\skipped}[1]{%
  \begingroup
    \color{gray!70}%
    \let\oldtextcolor\textcolor
    \renewcommand{\textcolor}[2][]{\oldtextcolor{gray!70}{##2}}%
    \let\color\relax
    #1%
  \endgroup
}

\usepackage{hyperref}
\hypersetup{
  colorlinks=true,
  linkcolor=Bittersweet,  
  citecolor=RoyalBlue,    
  urlcolor=teal
}
\usepackage[capitalise]{cleveref}

\begin{document}

\title{Nonlinear quantum noise polarization in macroscopic nonlinear QED}

\author{Arman Kashef}
\affiliation{Institut für Physik, Universität Rostock, Albert-Einstein-Straße 23-24, D-18059 Rostock, Germany}

\author{Marco Ornigotti}
\affiliation{Faculty of Engineering and Natural Sciences, 
Tampere University, Korkeakoulunkatu 7, FI-33720 Tampere, 
Finland}

\author{Stefan Scheel}
\affiliation{Institut für Physik, Universität Rostock, Albert-Einstein-Straße 23-24, D-18059 Rostock, Germany}
\email{stefan.scheel@uni-rostock.de}

\date{\today}

\begin{abstract}
    We construct a quantum theory of light in nonlinearly responding materials with dispersion 
    and absorption and, in particular, show the emergence of non-Gaussian noise associated with 
    nonlinear absorption processes. Our theory is based on a path-integral formulation of the 
    microscopic Huttner-Barnett model including cubic matter self-interaction required for 
    $\chi^{(2)}$ nonlinear processes. We show that the noise polarization emerges naturally from 
    a decoupling of the stochastic weights of the bath response which, in case of cubic 
    self-interactions, leads to a decoupling via Airy functions. As the theory is, by 
    construction, consistent with the nonlinear fluctuation-dissipation theorem, it constitutes 
    a viable quantum theory of light in nonlinear media.
\end{abstract}

\maketitle

\noindent\emph{Introduction} - Nonlinear optical processes are the basis for many applications in nanophotonics and photonics (quantum) information processing such as the generation of entangled photons \cite{entanglementReview,photonEnt1,agarwal}, squeezed light \cite{squeezing1,squeezing2,squeezing3}, frequency conversion \cite{Boyd}, lasing \cite{ultrafastNLO}, and integrated optics \cite{chipNLO}. The theory of perturbative nonlinear optics, starting from its inception in the 1960's \cite{Bloembergen}, assumes that the polarization field can be formally expanded in a series of increasing powers of the electric field, leading to a hierarchy of nonlinear optical processes involving nonlinear susceptibilities $\chi^{(n)}$ \cite{Boyd}. The causal nature of the optical response requires these susceptibilities to be complex functions of frequency, with their imaginary parts being associated with (nonlinear) absorption \cite{Bloembergen}. 

Quantum optics, on the other hand, is commonly based on the quantization of the electromagnetic field in vacuo or in strictly lossless media in order to ensure unitarity of the quantum evolution and the requirement for the electromagnetic field to obey quantum (Bose) statistics \cite{loudon, gerryKnight}. Combining causal nonlinear optical response and quantum theory poses a formidable challenge, as the unavoidable absorption destroys unitarity due to the coupling of the electromagnetic field with a dissipative environment.

A breakthrough in constructing a quantum theory of light in linearly responding, absorbing media was achieved by Huttner and Barnett \cite{HuttnerBarnett} who constructed a Hopfield-type model \cite{Hopfield} comprising the free electromagnetic field, coupled to a harmonic-oscillator field representing the matter polarization, which in turn is coupled to a continuum of harmonic oscillators responsible for the dissipation of energy from the electromagnetic field into the environment. This model, originally restricted to homogeneous and isotropic media, has been extended to include arbitrary linearly responding media \cite{philbin}, magnetodielectric materials \cite{magnetoHB}, epsilon-near-zero materials \cite{Bihb}, nonlinear optical materials \cite{praOrnigotti}, dual electromagnetism \cite{dualHB}, scattering problems \cite{ciattoni1,ciattoni2}, and has also been successfully applied to the propagation of nonclassical light through optical devices, or the theory of fluctuation-induced forces \cite{Acta}. 

Central to the consistency of the theory with requirements from open systems quantum theory and statistical physics in terms of the preservation of field commutators and the (linear) fluctuation-dissipation theorem is the appearance of a (linear) noise polarization field. This results in Langevin equations for the evolution of the electromagnetic field modes with Gaussian noise \cite{vogelWelsch,openQS}.

Several attempts at constructing a quantum theory of light in nonlinearly responding materials have been made \cite{drumondBook,Schmidt1998,Scheel2006,Crosse2010,sipeD,Lindel,saraviPRR}.  However, at present they all fall short of showing compliance with the nonlinear fluctuation-dissipation theorem \cite{HeinzWang}, as they do not allow for nonlinear absorption with an associated nonlinear noise polarization. A nonlinear quantum optical model capable of accounting for the required quantum noise profile is, however, necessary to benchmark and optimise the performance of photonic systems working at and below the quantum noise limit and would constitute an essential cornerstone in the design of quantum technologies based on nonlinear optics, such as squeezed light generation, quantum sensing, metrology, and frequency combs.

Recently, the Huttner-Barnett model has been extended to include anharmonic matter oscillators which result in $\chi^{(3)}$ nonlinear susceptibilities that obey Miller's rule \cite{Kashef26} based on a path-integral formulation \cite{Bechler, bookOrnigotti} of the model. The advantage of this approach is that it allows for a diagrammatic expansion of the dressed photon propagators and a systematic construction of higher-order nonlinear interactions. The starting point of this approach is an effective action including current sources for the electromagnetic and the matter polarization fields that, by differentiating the Gaussian part of the action integral with respect to the source fields, yield to a systematic expansion of the nonlinear effective action.

In this Letter, we extend this path-integral model to a Schwinger-Keldysh formalism that straightforwardly allows to construct retarded/advanced propagators that are required to show consistency with nonlinear fluctutation-dissipation theorems \cite{HeinzWang}. Within this formalism, we show how the noise polarization field naturally emerges from the theory and that, in order to fulfill the nonlinear fluctuation-dissipation theorem, 
its nonlinear part must necessarily include non-Gaussian noise. Our results allow a direct derivation and characterisation of the nonlinear quantum noise, allowing a realistic estimate of the performances of nonlinear photonic systems and exotic quantum states of light.

\noindent\emph{Schwinger-Keldysh form of the Huttner-Barnett model --} 
To begin, we consider a Lagrangian density, comprising
the free (transverse) electromagnetic field in the Coulomb gauge ($\nabla\cdot\mathbf{A}=0$), a harmonic-oscillator field $\mathbf{X}$ representing the matter degrees of freedom, a continuum of harmonic oscillators $\mathbf{Y}_\nu$ modelling the bath, and bilinear couplings between them. The total Lagrangian density is thus
\begin{equation}\label{eq:Ltot}
    \mathcal{L} = \mathcal{L}_{\text{EM}} + \mathcal{L}_{\text{mat}} + \mathcal{L}_{\text{bath}} + \mathcal{L}_{\text{EM-mat}} + \mathcal{L}_{\text{mat-bath}}\,,
\end{equation}
with
\begin{align}
    \mathcal{L}_{\text{EM}} &= \frac{1}{2}\left[\epsilon_0\,(\partial_t \mathbf{A})^2 - \frac{1}{\mu_0}\,(\nabla\times\mathbf{A})^2\right]\,,\label{eq:LEM}\\
    \mathcal{L}_{\text{mat}} &= \frac{\rho}{2}\left[(\partial_t \mathbf{X})^2 - \omega_0^2\,\mathbf{X}^2\right]\,,\label{eq:Lmat}\\
    \mathcal{L}_{\text{bath}} &= \int_0^\infty\!d\nu\;\frac{\rho_\nu}{2}\left[(\partial_t \mathbf{Y}_\nu)^2 - \nu^2\,\mathbf{Y}_\nu^2\right]\,,\label{eq:Lbath}\\
    \mathcal{L}_{\text{EM-mat}} &= \alpha'\, \mathbf{X}\cdot(\partial_t\mathbf{A})\,,\label{eq:Lint1}\\
    \mathcal{L}_{\text{mat-bath}} &= \int_0^\infty\!d\nu\;v(\nu)\,\mathbf{X}\cdot(\partial_t\mathbf{Y}_\nu)\,.\label{eq:Lint2}
\end{align}
Here, $\rho$ and $\rho_\nu$ are the mass densities for the matter polarization and the reservoir oscillators, respectively, $v(\nu)$ is the matter-bath coupling, and $\alpha'$ the coupling of the electric field to the matter polarization.
The bare retarded propagator for a harmonic oscillator with natural frequency $\omega_0$ is
\begin{equation}
G^R(\omega)= \frac{1}{(\omega  + i0^+)^2 - \omega_0^2}\,,
\end{equation}
where the infinitesimal positive imaginary frequency shifts its pole into the lower half-plane to ensure causality \cite{srednicki}. In the Schwinger-Keldysh formalism, the retarded and advanced propagators arise as specific combinations of the four possible contour-oriented correlators. The associated doubling of the integration contour induces a doubling of the fields, $\phi \to (\phi_+,\phi_-)$, corresponding to the forward and backward integration branches \cite{kamenev}. The generating functional takes then the form
\[
\mathcal{Z}= \int \mathcal{D}\phi_\pm\, e^{iS[\phi_+] - iS[\phi_-]},
\]
where in our case $\phi \in \{\mathbf{A}, \mathbf{X}, \mathbf{Y}_\nu\}$. 
The Keldysh rotation,
$\phi_{c}= \tfrac{1}{2}(\phi_+ + \phi_-)$ and $\phi_{q}= \phi_+ - \phi_-$, rearranges the fields into classical ($c$) and fluctuation ($q$) parts. 
Unitarity ($\mathcal{Z}=1$) implies $S_{\mathrm{eff}}[\phi_{q}=0]=0$ in which the identical histories carry no phase. As a result, setting $\phi_{q}=0$ yields the classical dynamics, whereas a nonzero $\phi_{q}$ captures both quantum and statistical fluctuations. After Keldysh rotation, any quadratic bosonic theory takes on a universal matrix structure, and the inverse Green’s function is triangular such that its action becomes
\begin{equation}\label{eq:S0_keldysh}
    S_0[\phi_{c} , \phi_{q}]=\frac{1}{2} \int\!\frac{d \omega}{2\pi}\;
    \bm{\Phi} ^\top(-\omega)
    \underbrace{\begin{pmatrix}
        0 & [G^{-1}]^{A} \\[4pt]
        [G^{-1}]^{R} & [G^{-1}]^{K}
    \end{pmatrix}}_{\displaystyle \mathbf{G}^{-1}(\omega)}
    \bm{\Phi}(\omega) \,,
\end{equation}
where $\bm{\Phi}^\top= \begin{pmatrix}\phi_{c} & \phi_{q}\end{pmatrix}$, and $[G^{-1}]^{R,A,K}$ are the inverse propagator kernels corresponding to the retarded (R), advanced (A), and Keldysh (K) propagators \cite{kamenev}. Integrating out the bath degrees of freedom $\mathbf{Y}_\nu$ results in the dressed-matter propagator 
\begin{equation}
G_X^R(\omega)= \frac{1}{\rho(\omega^2 - \omega_0^2) - \Sigma_{\text{bath}}^R(\omega)}
\end{equation}
for the polarization field $\mathbf{X}$ with the bath self-energy
\begin{equation}
\Sigma_{\text{bath}}^R(\omega)= \omega^2\int_0^\infty d\nu\,\frac{v(\nu)^2}{\rho_\nu}\,\frac{1}{(\omega + i0^+)^2 - \nu^2 }, 
\end{equation}
As a consequence of the Kubo–Martin–Schwinger (KMS) condition, 
the Keldysh self-energy reads
\begin{equation}
\Sigma_{\text{bath}}^K(\omega)= \coth\left(\frac{\omega}{2T}\right)\left[\Sigma_{\text{bath}}^R(\omega) - \Sigma_{\text{bath}}^A(\omega)\right].
\end{equation}
The statistical properties of the dressed-matter field are then captured by the Keldysh component, 
$G_X^K(\omega)$, which satisfies (see Sec.~I of the Supplementary Material \cite{supplementary})
\begin{equation}
G_X^K(\omega) =\frac{1}{2}G_X^R(\omega) \Sigma_{\text{bath}}^K(\omega) G_X^A(\omega),
\end{equation}
where $G_X^A(\omega)= [G_X^R(\omega)]^*$. This relation demonstrates that the fluctuations of the 
matter field ($G_X^K$) are the result of the bath's stochastic noise ($\Sigma_{\text{bath}}^K$), filtered through the system's dynamical response ($G_X^{R/A}$).

\noindent\emph{Linear noise polarisation --} The effective action for the electromagnetic field coupled to the dressed-matter field, when written in the Keldysh basis $(c,q)$, contains terms linear in $\mathbf{X}_q$ which, after functional integration, yields a functional delta distribution through
\begin{eqnarray}
    \int \mathcal{D}[X_q] \exp\left( i \int \frac{d\omega}{2\pi} X_q \cdot \text{EOM} \right) \nonumber\\ \propto \delta\left[ G_X^R(\omega)^{-1} X_{c}(\omega) + \alpha(\omega) A_{c}(\omega) \right],
\end{eqnarray}
where $\alpha(\omega)=-i\omega \alpha'$, and \text{EOM} stands for the equations of motion of the dressed-matter field, i.e. the argument of the functional delta distribution. This selects those trajectories that satisfy the semiclassical equations of motion \cite{kamenev,altland_condensed_2010}. The introduction of the bath adds a quadratic term $-\tfrac{1}{4}X_q \Sigma_{\text{bath}}^K X_q$ to the effective action, representing a stochastic weight. 
In order to maintain the Lagrange multiplier interpretation, we introduce a stochastic force field $\bm{\xi}(\omega)$ via a Hubbard-Stratonovich transformation \cite{strato1,strato2,hubbard1}
\begin{eqnarray}\label{hubbardStrato}
&\exp \left( -\frac{1}{2} \int \frac{d\omega}{2\pi} \mathbf{X}_q(-\omega) \mathcal{N}_{bath}(\omega) \mathbf{X}_q(\omega) \right) \\
&\propto\int \mathcal{D}[\bm{\xi}] \exp \left( \int \frac{d\omega}{2\pi} \left[ -\frac{1}{2} \bm{\xi}(-\omega) \mathcal{N}_{bath}^{-1} \bm{\xi}(\omega) + i \bm{\xi}(-\omega) \mathbf{X}_q(\omega) \right] \right) \nonumber
\end{eqnarray}
where $\mathcal{N}_{bath}(\omega)=\frac{i}{2}\Sigma_{\text{bath}}^K(\omega)$.

The equation of motion for the dressed-matter field now reads
\begin{equation}
\label{eq:dressedmattersolution}
\mathbf{X}_{c}(\omega) =G_X^R(\omega) \left[ -\alpha(\omega) \mathbf{A}_{c}(\omega) -\bm{\xi}(\omega) \right]
\end{equation}
which constrains the dressed-matter field to follow a Langevin trajectory. Substituting Eq. \eqref{hubbardStrato} back into the effective action formally replaces the term $\propto\mathbf{X}_q^2$ by a linear coupling $i\bm{\xi}\mathbf{X}_q$, de facto mapping the quantum action into a Langevin framework, with the auxiliary Hubbard-Stratonovich field $\boldsymbol\xi$ playing the role of the noise polarization source, whose correlations are determined by the Keldysh self-energy as
\begin{equation}
\label{xiNoise}
    \langle \bm{\xi}(\omega) \bm{\xi}(\omega') \rangle =2\pi\mathcal{N}_{bath}(\omega) \bm{\delta}(\omega + \omega').
\end{equation}
This formal procedure demonstrates that the noise polarization originates from the bath's fluctuations and is mapped to the matter field coordinates through the self-energy $\Sigma^K$.

The macroscopic polarization field ($\mathbf{P}=-\alpha' \mathbf{X}_{c}$) is thus
\begin{eqnarray}
\mathbf{P}(\omega) &= -{\alpha'}^2 G_X^R(\omega) {\mathbf{E}(\omega)} + \alpha' G_X^R(\omega) \bm{\xi}(\omega)  \nonumber\\ &\equiv \chi^{(1)}(\omega) \mathbf{E}(\omega) + \mathbf{P}^{(1)}_{\xi}(\omega),
\end{eqnarray}
where the noise polarization $\mathbf{P}^{(1)}_{\xi}$ represents the intrinsic thermal fluctuations of the dipole moment. Its correlation properties are inherited directly from the bath via Eq. \eqref{xiNoise}, i.e.
\begin{equation}
\langle \mathbf{P}^{(1)}_{\xi}(\omega) \mathbf{P}^{(1)}_{\xi}(\omega') \rangle =2\pi  \alpha'^2 |G_X^R(\omega)|^2 \mathcal{N}_{bath} \bm{\delta}(\omega + \omega') .
\end{equation}
The positive definiteness of $i\Sigma_{\text{bath}}^K(\omega)$ allows for a Kolmogorov-Cholesky decomposition $\bm{\xi}=\mathbf{L}(\omega)\,\mathbf{w}(\omega)$,
where $\mathbf{w}(\omega)$ is a white noise source, and $\mathbf{L}$ is the functional square root of the Keldysh self-energy, $\frac{i}{2}\Sigma^K =\mathbf{L}\mathbf{L}^\dagger$, with
\begin{equation}
\mathbf{L}(\omega) =\sqrt{-\coth\left(\frac{ \omega}{2T}\right) \text{Im}[\Sigma_{\text{bath}}^R(\omega)]}.
\end{equation}

Thus, in the Schwinger-Keldysh framework of the Huttner-Barnett model, the decomposition of the macroscopic polarization field into a reactive component involving the linear susceptibility $\chi^{(1)}$ of the model, and a noise polarization field by means of a Hubbard-Stratonovich transformation, is natural. Note also that the noise polarization field in the linear Huttner-Barnett model is represented by Gaussian noise which will no longer be true for nonlinear extensions as we will now show.

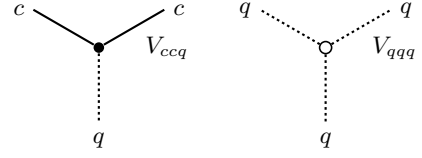
\begin{figure}
\begin{center}
\begin{tikzpicture}[scale=1]
\node[Vrra] (v1) at (0,0) {};
\draw[ext r] (v1) -- ++(150:1) node[left] {$c$};
\draw[ext r] (v1) -- ++(30:1) node[right] {$c$};
\draw[ext a] (v1) -- ++(270:1) node[below] {$q$};
\node[right=0.5cm] at (v1) {$V_{ccq}$};
 
\node[Vaaa] (v2) at (3,0) {};
\draw[ext a] (v2) -- ++(150:1) node[left] {$q$};
\draw[ext a] (v2) -- ++(30:1) node[right] {$q$};
\draw[ext a] (v2) -- ++(270:1) node[below] {$q$};
\node[right=0.5cm] at (v2) {$V_{qqq}$};
\end{tikzpicture}
\end{center}

\caption{Interaction vertices of the $X^3$ theory.}
\label{fig:X3vertices}
\end{figure}

\noindent\emph{Nonlinear noise polarisation --} We now employ the framework developed above 
to construct noise polarization fields for theories with a cubic self-interaction of the matter field. This implies an anharmonic potential for the matter polarization which, in this specific case, gives rise to a $\chi^{(2)}$ nonlinear susceptibility. In order to simplify the presentation, we treat all fields as scalar fields and leave the tensorial theory for future work. With this constraint, we add a cubic self-interaction to the Lagrangian as
\begin{equation}
    \mathcal{L} \;\longrightarrow\,\mathcal{L}+\mathcal{L}_{\rm int},
    \qquad \mathcal{L}_{\rm int} =-\frac{\lambda}{3!}\,X^3,
    \label{eq:Lint}
\end{equation}
with $\lambda\in\mathbb{R}$ being the nonlinear coupling constant. Doubling the field into forward and backward branches yields for the interaction term
\begin{equation}
    S_{\rm int} =-\frac{\lambda}{3!}\!\int\!d^4x\,\bigl[X_+^3-X_-^3\bigr],
\label{eq:Sint-pm}
\end{equation}
and, after Keldysh rotation, $X_\pm =X_c\pm\tfrac{1}{2}X_q$,
\begin{equation}
    S_{\rm int}[X_c,X_q]
    =-\frac{\lambda}{2}\!\int\!d^4x\,X_c^2\,X_q
      \;-\;\frac{\lambda}{24}\!\int\!d^4x\,X_q^3.
    \label{eq:Sint-clq}
\end{equation}
The two resulting vertices are shown in Fig.~\ref{fig:X3vertices}. The classical vertex $V_{ccq}\propto X_c^2 X_q$ survives in the semiclassical limit $X_q\mapsto0$ where it produces the nonlinear term $\propto\tfrac{\lambda}{2}X_c^2$ in the classical equations of motion. The remaining vertex $V_{qqq}\propto X_q^3$ does not affect the classical equations of motion but contributes to fluctuation vertices and to loop corrections. Other vertices do not appear, as
every interaction vertex must have an odd number of quantum legs.

After integrating out the bath, the effective action beyond the linear order in  $X_q$ takes the form
\begin{eqnarray}
    \label{eq:SXa}
    S_{X_q}
   &=&\int\!\frac{d\omega}{2\pi}\,
        \frac{i}{2}\,X_q(-\omega)\,\mathcal{N}_{\rm bath}(\omega)\,X_q(\omega) \\ &&
      - \frac{\lambda}{24}\!\int\!\!
        \frac{d\omega_1 d\omega_2 d\omega_3}{(2\pi)^3}\,(2\pi)\delta\!
        \left(\textstyle\sum_i\omega_i\right)
        X_q(\omega_1)X_q(\omega_2)X_q(\omega_3), \nonumber
\end{eqnarray}
The first term in Eq.~(\ref{eq:SXa}) is again quadratic in $X_q$ and can be treated by a Hubbard-Stratonovich transformation as shown above, resulting in a Gaussian noise field $\xi_G(\omega)$ with correlation given by Eq.~\eqref{xiNoise}.
The second term in Eq.~\eqref{eq:SXa}, however, is cubic in $X_q$ and cannot be removed by a linear shift of $\xi_G$. The decoupling of the cubic term thus requires a non-Gaussian auxiliary noise field. 

The cubic vertex is decoupled by the Airy Fourier pair
\begin{equation}
\int d\eta\,\frac{1}{\beta}\operatorname{Ai}\left(\frac{\eta}{\beta}\right)e^{i\eta X}
= e^{-i\beta^3 X^3/3}.
\end{equation}
Discretising time into slices $\delta t$ and rescaling the auxiliary field as
$\tilde\eta_n=\delta t\,\eta_n$ so that the source becomes
$\sum_n\tilde\eta_n X_{q,n}$, the single-slice Airy identity with
$\beta=(\lambda\delta t/8)^{1/3}$ reads
\begin{equation}
\int d\tilde\eta\,\frac{1}{\beta}\operatorname{Ai}\left(\frac{\tilde\eta}{\beta}\right)
e^{i\tilde\eta X}
= e^{-i\lambda\delta t X^3/24}.
\end{equation}
Combining independent time slices provides the exact finite-dimensional decoupling
\begin{equation}
\int d^N\tilde\eta\prod_n\frac{1}{\beta}\operatorname{Ai}\left(\frac{\tilde\eta_n}{\beta}\right)
e^{i\sum_n\tilde\eta_n X_{q,n}}
= e^{-i\lambda\delta t\sum_n X_{q,n}^3/24}.
\end{equation}
Requiring the logarithmic cumulant expansion of the weight to reproduce the cubic term forces all cumulants to vanish except for
\begin{equation}
\langle\eta(t_1)\eta(t_2)\eta(t_3)\rangle_c
= \frac{\lambda}{4}\delta(t_1-t_2)\delta(t_2-t_3).
\end{equation}
Thus, the auxiliary field represents white non-Gaussian noise with a single
delta-correlated third cumulant (for a full derivation, see Sec.~II of the Supplementary Material \cite{supplementary}.

Combining the Gaussian decoupling of the bath kernel with the Airy decoupling of the quantum vertex leads to the sought form of the noise force 
\begin{equation}
    \xi(t)
   =\underbrace{\xi_G(t)}_{\substack{\text{Gaussian,}\\
        \text{linear FDT}}}
      \;+\;
      \underbrace{\eta(t)}_{\substack{\text{non-Gaussian,}\\
        \text{from $V_{qqq}$}}}.
    \label{eq:xiP-decomposition}
\end{equation}
The semiclassical equations of motion then follow from the saddle-point condition 
$\delta S / \delta X_q = 0$ \cite{Kamenev01052009}, and read
\begin{equation}
\begin{aligned}
    \bigl[G_X^{R}&(\omega)\bigr]^{-1} X_c(\omega)
    = -\alpha(\omega)\,A_c(\omega) - \xi_G(\omega) \\[4pt]
    &\
      - \frac{\lambda}{24}\,
        \left(\tilde{\xi}_G
        \ast
        \tilde{\xi}_G\right)(\omega)
      +\frac{\lambda}{2}\,
        \bigl(X_c \ast X_c\bigr)(\omega),
\end{aligned}
\label{eq:Langevin-full}
\end{equation}
with $\tilde{\xi}_G(\omega) \;\equiv\; {\xi_G(\omega)}/{\mathcal{N}_{\rm bath}(\omega)}$ and $\ast$ representing convolution. The first line in Eq.~(\ref{eq:Langevin-full}) is the linear retarded response with the Gaussian noise source. The total noise takes the form of
Eq.~(\ref{eq:xiP-decomposition}), where the non-Gaussian part is given, to first order in 
$\lambda$, by the first term on the second line (see Sec.~III of the Supplementary Material \cite{supplementary} for details). The last term is the classical anharmonicity from $V_{ccq}$. 
If we now solve Eq.~(\ref{eq:Langevin-full}) perturbatively in $\lambda$ by writing $X_c=X^{(0)}_c + \,X^{(1)}_c + \cdots$ we retrieve, to lowest order,
\begin{equation}
    X^{(0)}_c(\omega)=G_X^R(\omega)\bigl[\mathcal{E}(\omega)-\xi_G(\omega)\bigr],
    \label{eq:X0}
\end{equation}
where $\mathcal{E}(\omega)\equiv-\alpha(\omega)A_c(\omega)=\alpha' E(\omega)$, which is the linear response, Eq.~(\ref{eq:dressedmattersolution}). The first-order correction has instead two contributions, one from each cubic vertex, namely
\begin{equation}
    X^{(1a)}_c(\omega) = \frac{\lambda}{2}\,G_X^R(\omega)\!
    \left( G_X^R\bigl[\mathcal{E}-\xi_G\bigr] \ast G_X^R\bigl[\mathcal{E}-\xi_G\bigr] \right)(\omega)
        \label{eq:X1a}
\end{equation}
resulting from the $V_{ccq}$ vertex, and
\begin{equation}
    X^{(1b)}_c(\omega) = -\frac{\lambda}{24}\,G_X^R(\omega)
      \,
        \left(\tilde{\xi}_G
        \ast
        \tilde{\xi}_G\right)(\omega),
        \label{eq:X1b}
\end{equation}
resulting from the $V_{qqq}$ vertex. A closer inspection to Eq. \eqref{eq:X1a} allows us to 
identify three distinct physical processes:
\begin{enumerate}
    \item[(i)] a term $\propto\mathcal{E}(\omega') \mathcal{E}(\omega-\omega')$, which we identify as a process involving a second-order susceptibility $\chi^{(2)}$,
    \item[(ii)] a term $\propto\mathcal{E}(\omega') \xi_G(\omega-\omega')$, which we identify as multiplicative noise,
    \item[(iii)] a term $\propto\xi_G(\omega') \xi_G(\omega-\omega')$ in (\ref{eq:X1a}) which, together with the contribution (\ref{eq:X1b}), is the source of the nonlinear noise polarization. 
\end{enumerate}
It should be noted that, had one naively replaced the classical electric field 
$\mathcal{E}(\omega)$ in the perturbative expansion of the macroscopic polarisation by its 
(linearly) fluctuating counterpart $\mathcal{E}(\omega)+\xi(\omega)$ \cite{Lindel}, one would 
have only been able to reproduce the noise structure in Eq. \eqref{eq:X1a}, arriving at the 
conclusion that the nonlinear noise is simply multiplicative and proportional to 
$\xi(\omega)\xi(\omega-\omega')$.
However, discarding the vertex contribution $X_c^{(1b)}$ breaks the quantum KMS symmetry of the Keldysh action~\cite{Sieberer2015}, and hence the quantum nonlinear fluctuation-dissipation theorem, as we will show in a separate publication.

Isolating the noise requires combining the $\xi_G\ast\xi_G$ term in Eq.~\eqref{eq:X1a} with 
Eq.~\eqref{eq:X1b}. The products $\xi_G\ast\xi_G$ are not purely stochastic: they carry non-zero 
mean values (tadpole shifts). On the other hand, the non-zero mean of $\tilde\xi_G*\tilde\xi_G$ 
is an artifact of the $\mathcal{O}(\lambda)$ representation of $\eta$ and must be removed.
Separating these mean values allows the deterministic (mean) parts to be rigorously subtracted,
leaving behind a genuinely zero-mean non-Gaussian noise polarization while leaving the nonlinear 
fluctuation--dissipation theorem intact. 
Having isolated the nonlinear noise as a zero-mean object (by keeping the $V_{qqq}$ vertex and 
subtracting their mean values), all contributions to the macroscopic polarization up to first order in $\lambda$ are accounted for. Combining the contributions to the noise as
\begin{equation}
\begin{aligned}
P_{(iii)+V_{qqq}}(\omega) &=P_{\xi\xi}(\omega)-\alpha'\big[X_c^{(1b)}(\omega)-\langle X_c^{(1b)}(\omega)\rangle\big]\\&= \langle P_{\xi\xi}^{\mathrm{tadpole}}(\omega)\rangle + P_{\xi\xi}^{\mathrm{NL}}(\omega),
\end{aligned}
\end{equation}
where $\langle P_{\xi\xi}^{\mathrm{tadpole}}(\omega)\rangle$ is the deterministic tadpole and 
$P_{\xi\xi}^{\mathrm{NL}}(\omega)$ is the normal-ordered, zero-mean nonlinear noise polarization 
satisfying $\langle P_{\xi\xi}^{\mathrm{NL}}(\omega)\rangle = 0$, the full polarization takes the 
form
\begin{widetext}
\begin{equation}
\begin{aligned}
P(\omega) &= \underbrace{
\underbrace{\chi^{(1)}(\omega)E(\omega)}_{\text{linear response}}+ \underbrace{\int \frac{d\omega'}{2\pi} \, \chi^{(2)}(\omega;\omega',\omega-\omega') E(\omega')E(\omega-\omega')}_{\text{process (i) }}
+ \underbrace{\langle P_{\xi\xi}^{\mathrm{tadpole}}(\omega)\rangle}_{\text{tadpole}}}_{\text{deterministic part }P_{\mathrm{det}}(\omega)} 
 + \underbrace{\underbrace{P_\xi^{(1)}(\omega)}_{\text{linear noise}}
+ \underbrace{P_{\mathrm{mult}}(\omega)}_{\text{process (ii)}}
+ \underbrace{P_{\xi\xi}^{\mathrm{NL}}(\omega)}_{\text{process (iii)+}V_{qqq}}
}_{\text{zero-mean noise part }P_\xi(\omega)}.
\end{aligned}
\end{equation}
\end{widetext}

\noindent\emph{Conclusions --} By utilizing the Schwinger-Keldysh path integral formulation of the nonlinear Huttner-Barnett model, we were able to consistently derive nonlinear noise polarizations that have hitherto been elusive. This places us in a situation in which we have a quantum-statistically consistent formulation of nonlinear macroscopic quantum electrodynamics with dispersion and absorption that, due to its consistency with the nonlinear fluctuation-dissipation theorem, allows one to investigate the influence of noise on nonlinear optical processes such as spontaneous parametric down-conversion or squeezed-light generation. We have shown that, by a suitable Hubbard-Stratonovich transformation of the stochastic weights, the quantum action is transformed into a Langevin framework in which one can identify the relevant noise sources. While the linear Huttner-Barnett model leads to a well-known Gaussian noise, a cubic self-interaction of the matter fields requires an Airy decoupling that necessarily leads to a non-Gaussian noise. 

It is expected that a quartic self-interaction as required for $\chi^{(3)}$ Kerr-nonlinearities will require yet another decoupling formulation, but will ultimately results in a similar calculation with only two vertex types contributing to the equations of motion. We expect this theory to have wide-ranging measurable consequences, from imposing physical constraints on the purity of photon pairs generated in spontaneous parametric down-conversion to limits on the generation of squeezing in nonlinear crystals.

\acknowledgments
\noindent\emph{Acknowledgments --} This work is funded by the Deutsche Forschungsgemeinschaft (DFG, German Research Foundation) through IRTG 2676/1 ‘Imaging of Quantum Systems’, project no. 437567992. M. O. acknowledges financial support from the Research Council of Finland Photonics Research and Innovation Flagship (PREIN: decision No. 320165).

\bibliographystyle{apsrev4-1}
\bibliography{Journal/biblio.bib}

\clearpage
\newgeometry{margin=3 cm}
\onecolumngrid
\setcounter{equation}{0}
\setcounter{figure}{0}
\setcounter{table}{0}
\setcounter{page}{1}
\setcounter{section}{0}
\setcounter{secnumdepth}{3}
\thispagestyle{empty}

\renewcommand{\theequation}{\arabic{equation}}
\renewcommand{\thefigure}{\arabic{figure}}
\renewcommand{\thetable}{\arabic{table}}
\renewcommand{\thesection}{\Roman{section}}
\renewcommand{\thesubsection}{\thesection.\Alph{subsection}}
\patchcmd{\section}{\centering}{\raggedright}{}{}
\makeatletter
\def\@hangfrom@section#1#2#3{\@hangfrom{#1#2}#3}%
\def\@hangfroms@section#1#2{#1#2}%
\makeatother

\titleformat{\section}
  {\normalfont\normalsize\bfseries}{\thesection}{1em}{}
\titleformat{\subsection}
  {\normalfont\normalsize\bfseries}{\thesubsection}{1em}{}

\makeatletter
\setcounter{NAT@ctr}{0}
\renewcommand{\bibnumfmt}[1]{[#1]}
\renewcommand{\citenumfont}[1]{#1}
\makeatother

\begin{center}
  {\large\bfseries Supplementary Material for ``Nonlinear quantum noise polarization in macroscopic nonlinear QED''}\\[3em]
\end{center}
In this Supplementary Material we provide detailed derivations and additional results supporting
the main text.
\section{Bath-dominated Keldysh--Dyson relation for an open oscillator}
We work in the average–difference (Keldysh) basis on the closed-time-path contour. The propagator  structure takes the form 
\begin{equation}
\hat{G} =
\begin{pmatrix}
G^K & G^R \\[4pt]
G^A & 0
\end{pmatrix},   \qquad
 \hat{G}^{-1} =
\begin{pmatrix}
0 & [G^A]^{-1} \\[4pt]
[G^R]^{-1} & [G^{-1}]^K
\end{pmatrix}.
\end{equation}
The $(2,2)$ component of the inverse propagator is a derived quantity given by the kinematic identity
\begin{equation}
    [G^{-1}]^K = -[G^R]^{-1} G^K [G^A]^{-1},
    \label{eq:ki-id}
\end{equation}
which holds for the bare and the dressed propagator alike. 
The dressed propagator $[G^{-1}]^K$ follows from the Dyson equation
\begin{equation}
    \hat{G}_X^{-1} = \hat{G}_{0,X}^{-1} - \hat{\Sigma}_{\rm bath},
    \label{eq:dys}
\end{equation}
in which the bare oscillator propagator is dressed by the bath self-energy. For a free oscillator of frequency $\omega_0$, the inverse retarded and advanced propagators are
\begin{equation}
[G_0^R(\omega)]^{-1} = (\omega + i0^+)^2 - \omega_0^2,
\qquad
[G_0^A(\omega)]^{-1} = (\omega - i0^+)^2 - \omega_0^2.
\end{equation}
The bare Keldysh propagator is parametrized by the oscillator's initial distribution
$F_0(\omega)$ as
\begin{equation}
G_0^K(\omega) =\frac{1}{2} \big( G_0^R(\omega) F_0(\omega) - F_0(\omega) G_0^A(\omega)\big).
\label{eq:dist}
\end{equation}
For a stationary scalar oscillator, $F_0$ is a function of frequency alone and commutes with $G_0^R$ and $G_0^A$. Substituting \eqref{eq:dist} into the algebraic identity \eqref{eq:ki-id}, this simplifies to
\begin{equation}
[G_{0,X}^{-1}]^K =\frac{1}{2} F_0(\omega) \Big( [G_0^R(\omega)]^{-1} - [G_0^A(\omega)]^{-1} \Big).
\end{equation}
so that
$
[G_{0,X}^{-1}]^K = 2i0^+ \omega \, F_0(\omega).
\label{eq:mat-kel}
$
Inserting Eq.~\eqref{eq:mat-kel} into Eq.~\eqref{eq:dys} gives the dressed inverse Keldysh component
\begin{equation}
[G_X^{-1}]^K = 2i0^+ \omega \, F_0(\omega) - \frac{1}{2}\Sigma_{\rm bath}^K(\omega).
\end{equation}
The order of limits is now essential. The bare contribution is proportional to $i0^+$, while the bath self-energy $\Sigma_{\rm bath}^K$ is finite. Without the bath self-energy, the product $[G_0^R]^{-1} G_0^K [G_0^A]^{-1}$ is singular: the poles of $G_0^R$ and $G_0^A$ lie on the real axis where the delta function in $G_0^K$ is supported, and the limit is not well-defined. The regulator is what makes this product meaningful. Once the bath is included, $\mathrm{Im}\,\Sigma_{\rm bath}^R \neq 0$ shifts the poles of the dressed propagators off the real axis.
Removing the regulator
after the bath is included, the bare term drops out and
\begin{equation}
[G_X^{-1}]^K = -\frac{1}{2}\Sigma_{\rm bath}^K(\omega).
\end{equation}
 Comparing with the algebraic identity \cref{eq:ki-id}, we find
\begin{equation}
    -[G_X^R]^{-1} \, G_X^K \, [G_X^A]^{-1} = -\frac{1}{2}\Sigma_{\rm bath}^K.
\end{equation}
Multiplying on the left by $G_X^R$ and on the right by $G_X^A$, we obtain
\begin{equation}
    G_X^K(\omega) =\frac{1}{2} G_X^R(\omega) \, \Sigma_{\rm bath}^K(\omega) \, G_X^A(\omega).
\end{equation}
This is the stationary, bath-dominated Keldysh--Dyson relation \citeSM{stefanucci_nonequilibrium_2013}. The bath self-energy $\Sigma_{\rm bath}^K$ is the noise source, and the dressed Keldysh Green's function is that noise weighted by the system's causal response at both ends: the retarded propagator $G_X^R$ filters noise emitted at earlier times, and the advanced propagator $G_X^A$ filters noise emitted at later times. The initial distribution $F_0$ does not appear because the retarded propagation that would carry it decays in the long-time limit, provided the bath produces genuine damping, $\mathrm{Im}\,\Sigma_{\rm bath}^R \neq 0$.

\section{The Airy decoupling of the cubic quantum vertex:
An exact auxiliary-field weight for non-Gaussian noise} Consider a degree of freedom $X$ linearly coupled to a thermal bath and carrying a cubic self-interaction. On the closed-time-path contour the field is doubled, $X\to(X_+,X_-)$ and the Keldysh rotation $X_r=\tfrac12(X_++X_-)$, $X_a=X_+-X_-$ separates the physical mean field from the fluctuation field. After eliminating the bath, the part of the action that depends on $X_a$ has the structure
\begin{equation}
S_{X_a}
= \int\frac{\mathrm{d}\omega}{2\pi}\, \frac{i}{2}\,X_a(-\omega)\,\mathcal{N}(\omega)\,X_a(\omega)
- g\!\int\!\mathrm{d} t\,X_a^3(t),
\label{eq:Sphia}
\end{equation}
where $N(\omega)\ge 0$ is the real, even noise kernel fixed by the bath spectral density and the KMS condition and $g\in\mathbb{R}$ is the cubic coupling. A cubic term $-(\lambda/3!)X^3$ in the underlying Lagrangian produces, after contour doubling and rotation, both a classical vertex 
$-(\lambda/2)X_r^2 X_a$ and the all-fluctuation vertex in \cref{eq:Sphia}. Only the latter concerns us, and $g$ stands for its coefficient (e.g.\ $g=\lambda/24$ in that normalization). 

The quadratic term in \cref{eq:Sphia} is decoupled by the standard Gaussian HS transformation, which introduces an additive Gaussian noise $\xi^{(G)}$ with $\langle\xi^{(G)}(\omega)\xi^{(G)}(\omega')\rangle = 2\pi\,\delta(\omega+\omega')\,\mathcal{N}(\omega)$ and renders the action linear in 
$X_a$. The Langevin representation rests on this linearity, because $\int\!\mathcal{D}X_a$ then yields a $\delta$-functional imposing the equation of motion. The cubic term spoils the linearity, and the question is whether an auxiliary field can restore it. We seek $\eta(t)$ and a weight $f[\eta]$ such that
\begin{equation}
\exp\Big(-i\,g\int\mathrm{d} t\,X_a^3(t)\Big)
= \int \mathcal{D}\eta\; f[\eta]\;
\exp\Big(i\int \mathrm{d} t\,\eta(t)\,X_a(t)\Big).
\label{eq:star}
\end{equation}
The right-hand side is the characteristic functional of the stochastic process $\eta$, evaluated at the source $X_a$. Constructing $f[\eta]$ is therefore equivalent to finding a process whose characteristic functional is the cubic phase on the left.

\paragraph{Exact decoupling}
One of the integral representations of the Airy function is the oscillatory integral \citeSM{abramowitz_handbook_2013,vallee_airy_2010}
\begin{equation}
\operatorname{Ai}(x) \equiv \frac{1}{2\pi}\int_{-\infty}^{\infty}\mathrm{d} u\,
\exp\!\Big(i\frac{u^3}{3} + i x u\Big),
\label{eq:Ai-def}
\end{equation}
that is convergent as an oscillatory (distributional) integral. Rotating the contour into sectors in which $\operatorname{Re}(iu^3)<0$ makes the integrand decay and makes $\operatorname{Ai}(x)$ a smooth real function. Equation~\eqref{eq:Ai-def} thus states that $\operatorname{Ai}(x)$ is the inverse Fourier transform of $e^{iu^3/3}$, so that
\begin{equation}
\int_{-\infty}^{\infty}\mathrm{d} x\,\operatorname{Ai}(x)\,e^{-ixu} = e^{iu^3/3}.
\label{eq:master}
\end{equation}
This is one of the structural counterparts of the Gaussian identity $\int \mathrm{d} x\,p_{_G}(x)\,e^{-ixu}=e^{-u^2/2}$. A Gaussian density has a quadratic-exponential Fourier transform while the Airy function has a cubic-exponential one. The Hubbard-Stratonovich decoupling uses the former 
to decouple a quadratic term, and the latter is exactly what decouples a cubic term. 

A rescaling fits \cref{eq:master} to an arbitrary cubic coefficient. Then for any $\beta>0$ and any $X\in\mathbb{R}$ it is straightforward to show that,
\begin{equation}
\int_{-\infty}^{\infty}\mathrm{d}\eta\,
\frac{1}{\beta}\,\operatorname{Ai}\Big(\frac{\eta}{\beta}\Big)\,e^{i\eta X}
= \exp\Big(-\frac{i\beta^3}{3}\,X^3\Big).
\label{eq:rescaled}
\end{equation}
In order to choose $\beta$ so that the right-hand side of \cref{eq:rescaled} becomes the coefficient that will appear on a single time slice, we discretise time into slices of width $\delta t$, with $X_{a,n}\equiv X_a(t_n)$. 
The Riemann sum $\int \mathrm{d} t\,X_a^3 \to \delta t\sum_n
X_{a,n}^3$ shows that the cubic action contributes
$-i\,g\,\delta t\,X_{a,n}^3$ on slice $n$. Matching the exponent of
\cref{eq:rescaled} to this fixes the per-slice Airy scale,
\begin{equation}
\;\beta(\delta t) = \big(3\,g\,\delta t\big)^{1/3}.\;
\label{eq:beta}
\end{equation}
With this scale, \cref{eq:rescaled} becomes the single-slice decoupling identity
\begin{equation}
\int_{-\infty}^{\infty}\!\mathrm{d}\eta\,
\frac{1}{\beta}\,\operatorname{Ai}\Big(\frac{\eta}{\beta}\Big)\,e^{i\eta X}
= e^{-i\,g\,\delta t\,X^3},
\label{eq:slice}
\end{equation}
an exact one-dimensional integral identity valid for every $\delta t>0$.

Now we combine subsequent slices. The source coupling $\int\mathrm{d} t\,\eta\,X_a$ discretises to $\delta t\sum_n \eta_nX_{a,n}$ and by absorbing the factor $\delta t$ into a slice-rescaled variable $\tilde\eta_n \equiv \delta t\,\eta_n$ gives $\sum_n \tilde\eta_nX_{a,n}$. 
Writing \cref{eq:slice} with $\tilde\eta_n$ as integration variable and $X_{a,n}$ as its conjugate source, and taking the product over the $N$ independent slices yields
\begin{equation}
\int\! \mathrm{d}^N\tilde\eta
\Big[\prod_{n}\frac{1}{\beta}\operatorname{Ai}\!\Big(\frac{\tilde\eta_n}{\beta}\Big)\Big]
e^{\,i\sum_n \tilde\eta_nX_{a,n}}
= e^{-i\,g\,\delta t\sum_n X_{a,n}^3}.
\label{eq:discrete-star}
\end{equation}
This is an exact, finite-dimensional identity for every $\delta t>0$, obtained
solely from \cref{eq:master}. The right-hand side is the sliced form of the cubic
action in \cref{eq:star}. The left-hand side is the sliced form of the auxiliary
integral, with discrete weight
$$f_{\rm slice}=\prod_n\beta^{-1}\operatorname{Ai}(\tilde\eta_n/\beta).$$

Approaching the continuum limit involves covering a fixed interval $[0,T]$ with $N$ slices, $N\,\delta t=T$. Refining the mesh ($N\to\infty$) forces $\{\delta t, \beta(\delta t)\}\to0$. The two scales
simultaneous vanishing is ambiguous in perturbative treatments \citeSM{LinBuLei2024}. The coarse-graining time and the auxiliary scale are therefore not independent. They are locked by \cref{eq:beta}.

Ref.~\citeSM{LinBuLei2024} identifies its non-Gaussian weight functions as non-positive-definite but does not characterize their status as continuum objects, since the construction is deliberately kept at finite $\delta t$. The present work supplies this missing piece for the cubic vertex. It constructs the $N\to\infty$ limit explicitly and identifies the resulting object as a continuous linear functional on test fields. That is fixed by its characteristic functional, and shows why no measure support it.

We emphasize that $\mathcal{D}\eta\,f[\eta]$ is a formal bookkeeping device, not a measure on field space. The construction is modeled on the Gelfand--Vilenkin formulation of generalized stochastic
processes \citeSM{gelfand_generalized_1976}, but the
cubic-phase characteristic functional fails positive-definiteness, so no
probability measure stands behind it. Formally, $\eta$ is a continuous
functional on smooth test fields $X_a$, specified by its characteristic
functional\begin{equation}
L[X_a]\equiv\Big\langle e^{\,i\int\mathrm{d} t\,\eta X_a}\Big\rangle,
\label{eq:char-functional}
\end{equation}
which we define as the $N\to\infty$ limit of the discrete left-hand side of
\cref{eq:discrete-star},
\begin{align}
L[X_a]
&\equiv
\lim_{N\to\infty}
\int \mathrm{d}^N\tilde\eta
\Big[\prod_{n}\frac{1}{\beta(T/N)}
\operatorname{Ai}\Big(\frac{\tilde\eta_n}{\beta(T/N)}\Big)\Big]
e^{\,i\sum_n \tilde\eta_nX_{a,n}}.
\label{eq:cont-def}
\end{align}

Collecting these results yields the final continuum identity.
The right-hand side of \cref{eq:discrete-star} converges to $\exp(-ig\int\!\mathrm{d} t\,X_a^3)$. The left-hand side has no pointwise continuum density, yet \cref{eq:discrete-star} is an exact equality for every finite $N$. The continuum left-hand side is well defined as a functional of the smooth test field $X_a$, not as a pointwise density. Combining \cref{eq:discrete-star,eq:cont-def} gives the continuum identity \cref{eq:star} that now has to be understood as an equality of characteristic functionals on the test-field space.

\paragraph{Cumulant structure}
The generating functional of $\eta$ is, by \cref{eq:star},
\begin{equation}
Z[X_a] = L[X_a] = \exp\Big(-i\,g\!\int\!\mathrm{d} t\,X_a^3(t)\Big),
\label{eq:Z}
\end{equation}
so the connected cumulants follow by functional differentiation,
\begin{equation}
\langle{\eta(t_1)\cdots\eta(t_n)}\rangle_c
= (-i)^n\,
\frac{\delta^n \ln Z}{\delta X_a(t_1)\cdots\delta X_a(t_n)}\bigg|_{X_a=0}.
\end{equation}
Since $\ln Z = -ig\int\!\mathrm{d} t\,X_a^3$ is exactly cubic, all derivatives of order $n\neq 3$ vanish at $X_a=0$, while the third cumumant gives
\begin{equation}\langle
{\eta(t_1)\eta(t_2)\eta(t_3)}\rangle_c= 6\,g\,\delta(t_1-t_2)\,\delta(t_2-t_3).
\label{eq:kappa3}
\end{equation}
Thus, $\eta$ is a white non-Gaussian noise with vanishing mean and variance, a
single delta-correlated third cumulant, and no higher cumulants. In the language of infinitely divisible laws, $\eta$ is the formal
$\alpha=3$ analogue of a stable process. However, stable probability laws exist only for $0 < \alpha \leq 2$. Therefore, although the observed scaling is consistent with the formal relation
$
\frac{1}{\alpha} = \frac{1}{3},
$
it cannot correspond to a stable probability law with \(\alpha = 3\) \citeSM{Sato1999}.

\section{From cubic vertex to non-Gaussian Langevin noise: A cumulant-matching derivation}
 The $X_q$-only part of the action defines the noise generating functional,
\begin{equation}
S_{\text{noise}}[X_q]
= \frac{i}{2}{\mathcal{N}_{\text{bath}}} X_q^2 - \frac{\lambda}{24} X_q^3, \label{eq:app-Snoise}
\end{equation}
for which we introduce a stochastic field $\xi$, distributed according to
\begin{equation}
e^{i S_{\text{noise}}[X_q]}
=\int \mathcal{D}\xi \; P[\xi] \; e^{i \int \xi X_q}.
\label{eq:app-defnoise}
\end{equation}
The field $X_q$ is the source that probes the distribution of $\xi$. Each power of $X_q$ corresponds to a cumulant of the noise. The space of probability distributions is infinite-dimensional and the cumulants are a natural coordinate system on it. Expanding both sides of~\eqref{eq:app-defnoise} in powers of \(X_q\),
\begin{equation}
-\frac{1}{2}\,\mathcal{N}_{\text{bath}}\, X_q^2 - \frac{i\lambda}{24} X_q^3 = i \langle \xi \rangle X_q - \frac{1}{2} \langle \xi \xi \rangle_c X_q^2 - \frac{i}{6} \langle \xi \xi \xi \rangle_c X_q^3 + \cdots .
\end{equation}
Matching term by term yields
\begin{equation}
\langle \xi \rangle = 0, \qquad
\langle \xi \xi \rangle_c = \mathcal{N}_{\text{bath}}, \qquad
\langle \xi \xi \xi \rangle_c = \frac{\lambda}{4}.
\label{eq:app-cumulants}
\end{equation}
To represent $\xi$ explicitly, we start with a Gaussian field $\xi_G$ with
\begin{equation}
\langle \xi_G \xi_G \rangle_c = \mathcal{N}_{\text{bath}}, \qquad
\langle \xi_G \xi_G \xi_G \rangle_c = 0,
\label{eq:app-Gaussian}
\end{equation}
and define
\begin{equation}
\xi \equiv \xi_G + c \, \xi_G^2.
\label{eq:app-shift}
\end{equation}
Here $\xi_G^2$ is Wick-ordered, $\xi_G^2-\langle\xi_G^2\rangle$ so $\langle\xi\rangle=0$. This is the mean subtraction of Eq.~(33) in the main text.
Defining the weakly non-Gaussian field $\xi$ with $\xi_G$ being Gaussian and $c \ll 1$ a small nonlinear coupling, we can expand all correlators in powers of $c$. Because connected correlators of a Gaussian field vanish beyond the two-point function, the leading connected three-point function is linear in $c$ and, by using Wick's we find
\begin{equation}
\begin{aligned}
\langle \xi\xi\xi \rangle_c &=
\langle \xi^3 \rangle - 3 \langle \xi \rangle \langle \xi^2 \rangle
+\mathcal{O}(c^2)\\ &=6c\,\mathcal{N}_{\text{bath}}^2,
\end{aligned}
\label{eq:app-cum3}
\end{equation}

Setting this equal to $\lambda/4$ gives
\begin{equation}
c = \frac{\lambda}{24\,\mathcal{N}_{\text{bath}}^2}.
\label{eq:app-c}
\end{equation}
In frequency space, squaring the noise becomes a convolution over frequency pairs $\omega'$ and $\omega-\omega'$
\begin{equation}
\xi(\omega)=\xi_G(\omega)+\frac{\lambda}{24}
\int \frac{d\omega'}{2\pi}\frac{\xi_G(\omega') \xi_G(\omega - \omega')}
{\mathcal{N}_{\text{bath}}(\omega') \mathcal{N}_{\text{bath}}(\omega - \omega')},
\label{eq:app-xiN}
\end{equation}
\bibliographystyleSM{apsrev4-2}
\bibliographySM{Journal/sup_bib}
\end{document}